\global\def\draftcontrol{0}

   \def\versionno{Heisenberg in Nonconformal Quivers}
\catcode`\@=11

\expandafter\ifx\csname draftcontrol\endcsname\relax\global\def\draftcontrol{0}
\fi

{\count255=\time\divide\count255 by 60
\xdef\hourmin{\number\count255}
\multiply\count255 by-60\advance\count255 by\time
\xdef\hourmin{\hourmin:\ifnum\count255<10 0\fi\the\count255}}
\def\draftdate{\number\month/\number\day/\number\year\ \ \ \hourmin }

\newcommand\makepapertitle{\par
  \begingroup
    \renewcommand\thefootnote{\@fnsymbol\c@footnote}%
    \def\@makefnmark{\rlap{\@textsuperscript{\normalfont\@thefnmark}}}%
    \long\def\@makefntext##1{\parindent 1em\noindent
            \hb@xt@1.8em{%
                \hss\@textsuperscript{\normalfont\@thefnmark}}##1}%
     \newpage
     \global\@topnum\z@   % Prevents figures from going at top of page.
     \@makepapertitle
     \thispagestyle{empty}\@thanks
  \endgroup
  \setcounter{footnote}{0}%
  \global\let\thanks\relax
  \global\let\makepapertitle\relax
  \global\let\@makepapertitle\relax
  \global\let\@thanks\@empty
  \global\let\@author\@empty
  \global\let\@date\@empty
  \global\let\@title\@empty
  \global\let\title\relax
  \global\let\author\relax
  \global\let\date\relax
  \global\let\and\relax
  \def\version{\let\version\@version\@gobble}
}
\def\@makepapertitle{%
  \newpage
   \ifnum\draftcontrol=1 {}
   \version\versionno
   \vskip 3em%
   \else
   \hfill\hbox to 3cm {\parbox{4cm}{\@pubnum}\hss}%
   \vskip 3em%
   \fi
   \begin{center}%
   \let \footnote \thanks
     {\LARGE {\@title}}%
     \vskip 1.5em%
     {\normalsize%\large
       \lineskip .5em%
       \begin{tabular}[t]{c}%
         \@author
       \end{tabular}\par}%
     \vskip 1.5em%
     {\@bstract}%
     \end{center}%
     \vskip 1.5em
     \@date%
   \par
}

\gdef\@pubnum{}
\def\pubnum#1{%
  \gdef\@pubnum{#1}}

\gdef\@bstract{}
\def\Abstract#1{%
  \gdef\@bstract{%
   \parbox{\textwidth-0pc}{%
   \centerline{\bf Abstract}\penalty1000%
\noindent%\abstractfont \baselineskip=12pt
\renewcommand\baselinestretch{1.0}%
{#1}}}
}

\def\ps@paper{\let\@mkboth\@gobbletwo%
     \ifnum\draftcontrol=1
        \def\@oddfoot{\hbox to \textwidth{\tiny \versionno \hfil\tiny\draftdate}%
        \hskip -\textwidth \hbox to \textwidth{\hfil\rm\thepage\hfil}}%
     \else\def\@oddfoot{\hbox to \textwidth{\hfil\rm\thepage\hfil}}
     \fi
     \let\@evenfoot\@oddfoot
}
\def\@version#1{\ifnum\draftcontrol=1
\typeout{}\typeout{#1}\typeout{}
\vskip3mm\centerline{\hbox{\fbox{\normalsize{\tt DRAFT -- #1 -- }
                   {\draftdate}}}}\vskip3mm
\fi}
\let\version\@version
\long\def\eqlabel#1{\ifnum\draftcontrol=1
                    \tag@false  % there are some problems with multline without this
                    \tag*{(\theequation) \hbox to -0.2cm{\hspace{0cm}\small{#1}\hss}}
                    \refstepcounter{equation}
                    \edef\@currentlabel{\theequation}
                    \ltx@label{#1}          % use old LaTeX \label instead of new definition
                    \else
                    \label{#1}
                    \fi
                    }
\let\st@bibitem\@bibitem
\let\st@lbibitem\@lbibitem
\ifnum\draftcontrol=1
  \def\@bibitem#1{%
    \st@bibitem{#1}\a@@label{#1}\ignorespaces}
  \def\@lbibitem[#1]#2{%
    \st@lbibitem[#1]{#2}\a@@label{#2}\ignorespaces}
  \def\a@@label#1{%
    \gdef\a@lab{\smash{\normalfont\small#1}}
    \ifvmode
      \if@inlabel
        \global\setbox\@labels\hbox{%
          \llap{\a@lab\let\a@lab\relax
                \kern\@totalleftmargin\kern\marginparsep}%
          \box\@labels}%
      \fi
    \fi}
\fi
\documentclass[12pt,letterpaper]{article}
\usepackage{amsmath,amssymb,array,calc,rotating,epsfig,psfrag}
\usepackage[nosort]{cite}
\usepackage{xypic}
\usepackage{diagxy}
\usepackage{diagrams}
\ifnum\draftcontrol=1
\fi
\renewcommand\baselinestretch{1.25}
\renewcommand\section{\@startsection {section}{1}{\z@}%
                                   {-3.5ex \@plus -1ex \@minus -.2ex}%
                                   {2.3ex \@plus.2ex}%
                                   {\normalfont\large\bfseries}}
\renewcommand\subsection{\@startsection{subsection}{2}{\z@}%
                                   {-3.25ex\@plus -1ex \@minus -.2ex}%
                                   {1.5ex \@plus .2ex}%
                                   {\normalfont\normalsize\bfseries}}
\renewcommand\subsubsection{\@startsection{subsubsection}{3}{\z@}%
                                   {-3.25ex\@plus -1ex \@minus -.2ex}%
                                   {1.5ex \@plus .2ex}%
                                   {\normalfont\normalsize\it}}
\renewcommand\paragraph{\@startsection{paragraph}{4}{\z@}%
                                   {-3.25ex\@plus -1ex \@minus -.2ex}%
                                   {1.5ex \@plus .2ex}%
                                   {\normalfont\normalsize\bf}}

\def\revise#1       {\raisebox{-0em}{\rule{3pt}{1em}}%
                     \marginpar{\raisebox{.5em}{\vrule width3pt\
                     \vrule width0pt height 0pt depth0.5em
                     \hbox to 0cm{\hspace{0cm}{%
                     \parbox[t]{4em}{\raggedright\footnotesize{#1}}}\hss}}}}

\def\zet          {{\mathbb Z}}
\def\Id           {{\mathbb I}}

\def\del          {\partial}

\def\tr           {\mathop{\rm Tr}}

\def\half{{\frac12}}

\def\sqr#1#2{{\vcenter{\vbox{\hrule height.#2pt
 \hbox{\vrule width.#2pt height#1pt \kern#1pt
 \vrule width.#2pt}\hrule height.#2pt}}}}
\def\square{%
  \mathop{\mathchoice{\sqr{12}{15}}{\sqr{9}{12}}{\sqr{6.3}{9}}{\sqr{4.5}{9}}}}

\newcommand{\fft}[2]{{\frac{#1}{#2}}}
\newcommand{\ft}[2]{{\textstyle{\frac{#1}{#2}}}}

\def\a{\alpha}
\def\b{\beta}
\def\r{\rho}

\def\m{\mu}
\def\g{\gamma}

\def\n{\nu}
\def\bn{\bar{\nu}}
\def\bm{\bar{\mu}}

\newcommand{\Tr}{{\rm Tr\,}}
\catcode`\@=12

\usepackage{color}

\begin{document}

%%%%%%%%%%%%%%%%%%%%%%%%%%%%%%%%%%%%%%End of Alex Draft Mode%%%%%%%%%%%%%%%%%%%%

%%%%%%%%%%%%%%%%%%From Carlo's Style%%%%%%%%%%%%%%%%%%%%%%%%%%%%%%%%%%5

%\overfullrule=0pt
%\parskip=2pt
%\parindent=12pt
%\headheight=0in
%\headsep=0in
%\topmargin=0.50in
%\oddsidemargin=0in

%--------+---------+---------+---------+---------+---------+---------+
\newcommand{\be}{\begin{equation}}
\newcommand{\ee}{\end{equation}}
\newcommand{\beq}{\begin{equation}}
\newcommand{\eeq}{\end{equation}}
\newcommand{\ba}{\begin{eqnarray}}
\newcommand{\ea}{\end{eqnarray}}
\newcommand{\nn}{\nonumber}
%--------+---------+---------+---------+---------+---------+---------+

% Ofer's definitions
\def\vol{\bf vol}
\def\Vol{\bf Vol}
\def\del{{\partial}}
\def\vev#1{\left\langle #1 \right\rangle}
\def\cn{{\cal N}}
\def\co{{\cal O}}
%\newfont{\Bbb}{msbm10 scaled 1200}     %instead of eusb10
%\newcommand{\mathbb}[1]{\mbox{\Bbb #1}}
\def\IC{{\mathbb C}}
\def\IR{{\mathbb R}}
\def\IZ{{\mathbb Z}}
\def\RP{{\bf RP}}
\def\CP{{\bf CP}}
\def\Poincare{{Poincar\'e }}
\def\tr{{\rm tr}}
\def\tp{{\tilde \Phi}}
\def\Y{{\bf Y}}
\def\te{\theta}
\def\bX{\bf{X}}

\def\TL{\hfil$\displaystyle{##}$}
\def\TR{$\displaystyle{{}##}$\hfil}
\def\TC{\hfil$\displaystyle{##}$\hfil}
\def\TT{\hbox{##}}
\def\HLINE{\noalign{\vskip1\jot}\hline\noalign{\vskip1\jot}} %Only in
\def\seqalign#1#2{\vcenter{\openup1\jot
  \halign{\strut #1\cr #2 \cr}}}
\def\lbldef#1#2{\expandafter\gdef\csname #1\endcsname {#2}}
\def\eqn#1#2{\lbldef{#1}{(\ref{#1})}%
\begin{equation} #2 \label{#1} \end{equation}}
\def\eqalign#1{\vcenter{\openup1\jot
    \halign{\strut\span\TL & \span\TR\cr #1 \cr
   }}}
\def\eno#1{(\ref{#1})}
\def\href#1#2{#2}
\def\half{{1 \over 2}}

%--------+---------+---------+---------+---------+---------+---------+
%Hirosi's macros:
\def\ads{{\it AdS}}
\def\adsp{{\it AdS}$_{p+2}$}
\def\cft{{\it CFT}}

\newcommand{\ber}{\begin{eqnarray}}
\newcommand{\eer}{\end{eqnarray}}

\newcommand{\bea}{\begin{eqnarray}}
\newcommand{\eea}{\end{eqnarray}}

\newcommand{\beqar}{\begin{eqnarray}}
\newcommand{\cN}{{\cal N}}
\newcommand{\cO}{{\cal O}}
\newcommand{\cA}{{\cal A}}
\newcommand{\cT}{{\cal T}}
\newcommand{\cF}{{\cal F}}
\newcommand{\cC}{{\cal C}}
\newcommand{\caR}{{\cal R}}
\newcommand{\cW}{{\cal W}}
\newcommand{\eeqar}{\end{eqnarray}}
\newcommand{\lm}{\lambda}\newcommand{\Lm}{\Lambda}
\newcommand{\eps}{\epsilon}

%--------+---------+---------+---------+---------+---------+---------+

\newcommand{\nonu}{\nonumber}
\newcommand{\oh}{\displaystyle{\frac{1}{2}}}
\newcommand{\dsl}
  {\kern.06em\hbox{\raise.15ex\hbox{$/$}\kern-.56em\hbox{$\partial$}}}
\newcommand{\as}{\not\!\! A}
\newcommand{\ps}{\not\! p}
\newcommand{\ks}{\not\! k}
\newcommand{\D}{{\cal{D}}}
\newcommand{\dv}{d^2x}
\newcommand{\Z}{{\cal Z}}
\newcommand{\N}{{\cal N}}
\newcommand{\Dsl}{\not\!\! D}
\newcommand{\Bsl}{\not\!\! B}
\newcommand{\Psl}{\not\!\! P}
\newcommand{\eeqarr}{\end{eqnarray}}
\newcommand{\ZZ}{{\rm \kern 0.275em Z \kern -0.92em Z}\;}

%%%%%%%%%%%%%%%%%Some more definitions%%%%%%%%%%%%%%%%%%%5
\def\s{\sigma}
\def\a{\alpha}
\def\b{\beta}
\def\r{\rho}
\def\d{\delta}
\def\g{\gamma}
\def\G{\Gamma}
\def\ep{\epsilon}
%%%%% number equations by section %%%%%%%%
\makeatletter \@addtoreset{equation}{section} \makeatother
\renewcommand{\theequation}{\thesection.\arabic{equation}}
%%%%%%%%%%%%%%%%%%%%%%%%%%%%%%%%%%%%%%%

%%%%%%%%%%%%%%Leo's%%%%%%%%%%%%%%%%%%%%5
\def\be{\begin{equation}}
\def\ee{\end{equation}}
\def\bea{\begin{eqnarray}}
\def\eea{\end{eqnarray}}
\def\m{\mu}
\def\n{\nu}
\def\g{\gamma}
\def\p{\phi}
\def\L{\Lambda}
\def \W{{\cal W}}
\def\bn{\bar{\nu}}
\def\bm{\bar{\mu}}
\def\bw{\bar{w}}
\def\ba{\bar{\alpha}}
\def\bb{\bar{\beta}}

\begin{titlepage}

\version\versionno

\leftline{\tt hep-th/06mmnnn}

\vskip -.8cm

\rightline{\small{\tt MCTP-06-03}}

\vskip 1.7 cm

\centerline{\bf \Large Central Extensions of Finite Heisenberg Groups}
\vskip .6cm
\centerline{\bf \Large in Cascading Quiver Gauge Theories}

%\vskip .6cm
%\centerline{\bf \Large  in Sasaki-Einstein Superconformal Quiver Theories}
\vskip 1cm
\vskip 1cm
{\large }
\vskip 1cm

\centerline{\large Benjamin A. Burrington, James T. Liu, and Leopoldo A. Pando
Zayas }

\vskip 1cm
\centerline{\it Michigan Center for Theoretical
Physics}
\centerline{ \it Randall Laboratory of Physics, The University of
Michigan}
\centerline{\it Ann Arbor, MI 48109-1040}

\vspace{1cm}

\begin{abstract}
Many conformal quiver gauge theories admit nonconformal generalizations. These generalizations
change the rank of some of the gauge groups in a consistent way, inducing a running
in the gauge couplings. We find a group of discrete
transformation that acts on a large class of these theories. These transformations form
a central extension of the Heisenberg group, generalizing the Heisenberg group
of the conformal case, when all gauge groups have the same rank.
In the AdS/CFT correspondence the nonconformal quiver gauge theory is dual to supergravity backgrounds with
both
five-form and three-form flux. A direct implication is that operators counting wrapped branes satisfy a central
extension of a finite Heisenberg group and therefore do not commute.

\end{abstract}

%\end{center}

%\noindent

\end{titlepage}

%\newpage

%--------+---------+---------+---------+---------+---------+---------+
%

%%%%%%%%%%%%%%%%%%%%%%%%%%%%%%%%%%%%%%%%%%%%%%%%%%%%%%%%%%%%%%%%%%%%%%%%%%%
\section{Introduction}
%%%%%%%%%%%%%%%%%%%%%%%%%%%%%%%%%%%%%%%%%%%%%%%%%%%%%%%%%%%%%%%%%%%%%%%%%%
A large class of quiver gauge theories admits the action of certain finite Heisenberg groups \cite{heisAlg}. In this paper we
continue our investigation of discrete symmetries in quiver gauge theories. In particular, we focus on nonconformal quivers.
This is a very interesting class of quiver gauge theories which is anomaly free and  nonconformal in the sense that the beta
functions corresponding to the gauge couplings are nonzero. This is a very interesting dynamical generalization which lead
to interesting QCD-like behavior as exemplified in the case of Klebanov-Strassler model \cite{ks} which displays confinement and chiral
symmetry breaking.

Let us recall the main result of \cite{heisAlg} which is a generalization of  \cite{grw}.  For a large class of
quiver gauge theories with gauge group $SU(N)^p$, there is a set of
discrete transformations $A, B$ and $C$ satisfying
\be
\eqlabel{heisen}
A^{q}=B^{q}=C^{q}=1, \qquad AB=BAC,
\ee
where $q$ is some integer
number which depends on the particular structure of the quiver.  These
transformations satisfy three important properties: (i) leave the
superpotential invariant, (ii) satisfy the anomaly cancelation for all
$SU(N)$ gauge groups, and (iii) the above group relations are true up to
elements in the center of the gauge group $SU(N)^p$, that is, up to gauge transformations.

The main result of this paper can then be formulated as follows. For generalizations of conformal quivers into nonconformal
quivers, that is,  we consider gauge theories with gauge group $\prod\limits_{i=1}^p SU(N+\alpha_i M)$ with $\alpha_i$ some
positive integers, we find a set of
discrete transformations $A, B,C$ and $D$ satisfying
\be
\eqlabel{centheisen}
A^q=B^q=C^q=D^q=1, \qquad AB=BAC, \quad AC=CAD,
\ee
here $q$ is the same integer as in the conformal case for general quivers. 
The conditions are the same as above, that is, invariance of the
superpotential, anomaly cancelation and the relations are true up to elements in the center of the gauge group.

An alternative way
of describing the above group is as a central extension of the Heisenberg group acting in the conformal case. Let us denote the
finite Heisenberg group acting in the conformal case and whose commutation relations are given
in (\ref{heisen}),  as ${\rm Heis}(\mathbb{Z}_q\times \mathbb{Z}_q)$, then the centrally extended group $H_q$ whose 
commutations relations are (\ref{centheisen}),
is defined via the short exact sequence:

\[\bfig
\morphism(0,200)|a|/{->}/<500,0>[\Id`{\zet_q};]
\morphism(500,200)|a|/{->}/<500,0>[{\zet_q }`{H_q};]
\morphism(1000,200)|a|/{->}/<500,0>[{H_q}`{{\rm Heis}(\mathbb{Z}_q\times \mathbb{Z}_q)};]
\morphism(1500,200)|a|/{->}/<500,0>[{{\rm Heis}(\mathbb{Z}_q\times \mathbb{Z}_q)}`\Id;]
\efig\]
where the $\mathbb{Z}_q$ factor is generated by the central element $D$ in (\ref{centheisen}). Interestingly, the central element 
in ${\rm Heis}(\mathbb{Z}_q\times \mathbb{Z}_q)$ which is denoted by $C$ in (\ref{heisen}) is no longer central as a element
of $H_q$ in (\ref{centheisen}). 
In section \ref{meat}  we will explicitly construct all the morphisms involved in the above sequence. Note that when $D$ is the identity one recovers the
Heisenberg group of the conformal limit. More precisely, the number $M$ in the gauge groups determines the structure of the
element $D$: when $M=0$ we have that $D=1$. Thus, the nontriviality of $D$ is directly related to the three-form flux which is
proportional to $M$. Alternatively, we can view $M$ as the number of fractional D5 branes in the string theory side.

To a large extent our investigation is motivated by ideas put forward by D. Belov, G. Moore and others suggesting that
D-brane charge in string theory with background RR flux is a noncommutative quantity \cite{kitp,dima}. In this respect our work
exploits the AdS/CFT correspondence \cite{agmoo} to learn about fundamental properties of D-branes. We also find the study of
discrete symmetries in quiver gauge theories interesting in its own right.

The organization of this note is as follows. In section \ref{review} we review the essential properties of quiver 
gauge theories that are further used in this paper. In section \ref{meat} we explicitly discuss various examples and give
some ideas of what a  general proof would entail. Section \ref{string} contains some comments on the implications for
the string theory description of these cascading quiver gauge theories. In section \ref{conclusions} we conclude with some observations
about the
limitations of our calculations.

%%%%%%%%%%%%%%%%%%%%%%%%%%%%%%%%%%%%%%%%%%%%%%%%%%%
\section{Generalities of Quiver Theories}\label{review}
%%%%%%%%%%%%%%%%%%%%%%%%%%%%%%%%%%%%%%%%%%%%%%%%%%%%%
Here we will review some of the general techniques of analysis
used in quiver field theories.  Our goal is to be self consistent and this section can clearly be skipped by 
readers who are familiar with the standard properties of quiver gauge theories. 

First, we will discuss the
role of the adjacency matrix in determining the ranks of the
gauge groups.  The adjacency matrix component $a_{ij}$ is defined
as the number of arrows pointing {\it from} the $i^{\rm th}$ node
{\it to} the $j^{\rm th}$ minus the number pointing {\it from} the $j^{\rm th}$ node
{\it to} the $i^{\rm th}$ node.  Thus, even though
there is an entry $0$ in the adjacency matrix, one may not conclude that
there are no arrows between the nodes.  For example, the conifold theory
\begin{center}
\includegraphics[width=.25\textwidth]{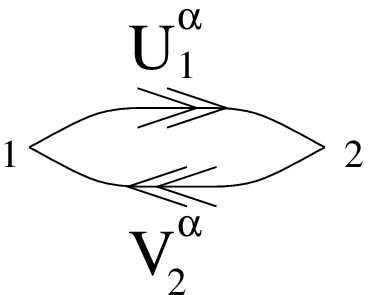},
\end{center}
has a two by two adjacency matrix with all entries being $0$.
A general adjacency matrix $\hat{a}$ is an antisymmetric matrix,
and has a certain number of zero eigenvectors that are important
for our purposes.
The adjacency matrix is a matrix of integers, and so we may always scale
its zero eigenvectors
to have integer components.  Any one of these integer valued zero
eigenvectors is a good assignment of gauge groups, assigning
the rank of the $i^{\rm th}$ $SU$ gauge group be the
$i^{\rm th}$ component
of that zero eigenvector.

This procedure is simply making sure that the
triangle anomaly cancels for any given node (gauge group)
of the quiver diagram.  Consider a node that denotes
a gauge group $SU(N)$.  Focusing on that node, the triangle
anomaly is proportional to $\Sigma \Tr(t_i^a t_i^b t_i^c)$, where
the sum runs over all other indices (we use the shorthand $i$)
that label fields, and
$t_i$ are the generators for the representation of this $i^{\rm th}$
field under the $SU(N)$.  This Casimir is zero
for the adjoint representation of an $SU$ group, and so
only the matter sector contributes.  The fundamental and
anti-fundamental representations of $SU$ contribute with opposite
sign.  The sum over other indices includes the
gauge indices from the other end of the arrow.  Hence, an arrow
pointing from a gauge group with gauge group
$SU(N')$ to the gauge group in question gets an additional factor
of $N'$ from the sum.  Therefore, to count the anomaly,
we can simply count the number
of arrows from a given gauge group to the group in question and
weight this arrow by the rank of the gauge group at the other end, where the sign
is given as $(+)$ for arrows pointing away, and $(-)$ for arrows pointing
towards the node in question.
This is precisely what $\hat{a}\overrightarrow{v}$ measures, the $i^{\rm th}$
entry being the anomaly at the $i^{\rm th}$ node, which we of course require
to be $0$.  This argument has nothing to do with supersymmetry, as the anomaly
cancels as long as the arrows represent the same type of fields (here, we are
considering that they are all chiral superfields).

%It is particularly useful to find a basis of zero eigenvectors such
%that all zero eigenvectors are integer weighted sums of these vectors.
%In such a basis, the full space of integer valued
%zero eigenvectors is ${\zet}^k$.

For illustrative purposes, we will use a toy example.
Consider the quiver diagram
\begin{center}
\includegraphics[width=.25\textwidth]{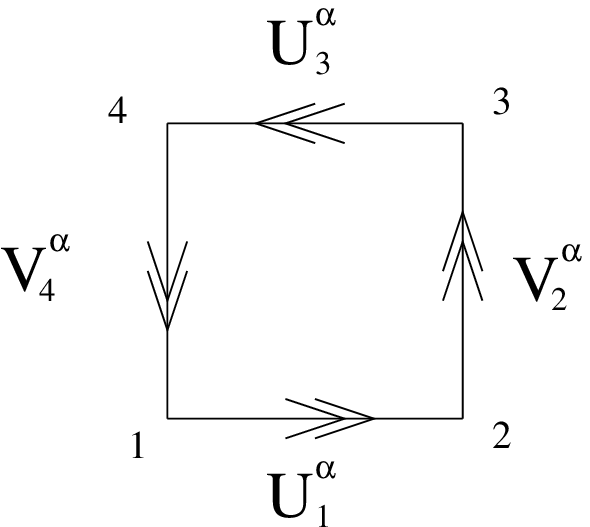}.
\end{center}
The adjacency matrix is
\be
\hat{a}=
\begin{pmatrix}
0 & 2 & 0 & -2 \\
-2 & 0 & 2 & 0 \\
0 & -2 & 0 & 2 \\
2 & 0& -2 & 0
\end{pmatrix}
\ee
There are two linearly independent solutions to the equation
$\hat{a} \overrightarrow{v}=0$ and they are
\be
\overrightarrow{v}_1=
\begin{pmatrix}
1 \\
1 \\
1 \\
1
\end{pmatrix}, \quad
\overrightarrow{v}_2=
\begin{pmatrix}
0 \\
1 \\
0 \\
1
\end{pmatrix}.
\ee
Any integer valued zero eigenvector can be expressed as
\be
\overrightarrow{v}_0=N\overrightarrow{v}_1+M\overrightarrow{v}_2
\ee
with $N$ and $M$ being integers.  Thus,
the assignment of gauge groups is as follows:
\begin{center}
\includegraphics[width=.25\textwidth]{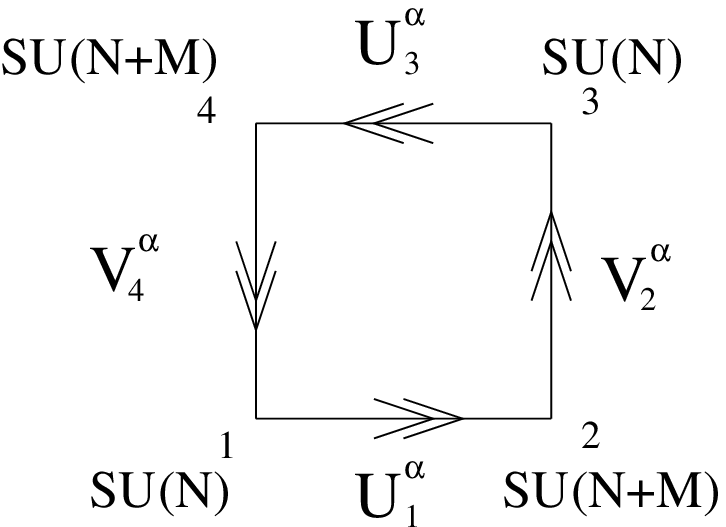}.
\end{center}
For these to be gauge groups,
we require that $N+M>1,\; N>1$ so that the assignments of
$SU(N)$ and $SU(N+M)$ as the gauge groups makes sense.

Now we go on to consider the $\beta$ functions for such
quivers.  For our purposes, we will always consider quivers
that have an assignment of gauge group ranks such that there
is a conformal case (the need for this assumption will
become clear in a moment).  These will correspond to vectors of the
type $\overrightarrow{v}_1$ with $1$ in every entry.

First, let us consider the $\beta$ functions associated
with the superpotential couplings.  As a short hand, we
will refer to these couplings as $h_i$.  Non renormalization
theorems simply give that a monomial term in a superpotential
with coupling $h_i$ has a beta function given in terms of the
anomalous dimension of the fields that enter the monomial,
\be
\beta_{h_i}=h_i(\mu)\left(-3+\Sigma \left(1+\ft12 \gamma_{ij} \right)\right)
\ee
where the anomalous dimension $\gamma_{ij}$ and sum refer to all fields present
in the superpotential monomial.  We use the two gauge groups under which the
field is charged to label the anomalous dimension.  In this discussion,
the anomalous dimension is defined through $dim(F_{ij})=1+\fft12\gamma_{ij}$ so that kinetic terms
are scale as $(1+\gamma_{i j})$ to leading order.  Knowing the $\gamma_{ij}$ is equivalent
to knowing the $R$ charge of the operator, $dim(\mathcal{O})=\frac32 R_{\mathcal{O}}$, in a
conformal theory.

The procedure for determining the $\beta_i$ for the gauge couplings
$g_i$ is much the same as considering the anomaly: they are determined
by considering the other other gauge groups in the diagram
as flavor symmetries.  In the general case for a node with gauge group $SU(N_i)$ and
gauge coupling $g_i$, the NSVZ beta function is
\be
\beta_i=-\frac{g_i^3}{16 \pi^2}\frac{3N_i-\Sigma N_j T_{r_{ij}}(1-\gamma_{ij})}{1-\frac{g_i^2 N_i}{8\pi^2}}.
\ee
$T_{r_{ij}}$ is the Casimir for the field charged under gauge groups
$i$ and $j$ given as $\Tr(t_{ij}^a t_{ij}^b)=T_{r_{ij}}\delta^{ab}$, and
the generators $t$ are for the $i^{\rm th}$ gauge group under investigation.
$\gamma_{ij}$ is the anomalous dimension of this field, and $N_j$ is the rank
of the gauge group at the node $j$.
%$N_{f_{ij}}$ is counted as the total number of arrows pointing {\it only into} the node from $j$
%to $i$, weighted
%by the gauge factors at the other end of the arrow and times $1/2$.  This is done so that our counting of
%the number of flavors, usually associated with a {\it pair} of chiral superfields, is the conventional
%counting.

The anomalous dimensions of the fields are found by taking the conformal case (the $v_1$ type
vector) and solving $\beta_i=\beta_{h_i}=0$.  Of course $a$ maximization is the real principle
that allows for solving for the anomalous dimensions, however here we will not concern ourselves
with the actual values, and only take that there is a solution that sets all the $\beta$ functions
to zero.  In the conformal cases we are considering, all nodes have rank $N$ gauge groups, and
so all terms present are all proportional to $N$.  Now, we assume that the other integer $M$ multiplying
the other zero eigenvector is small.  The assignment of $\gamma_{ij}$ can be seen to depend on $(M/N)^2$
and so if we work to leading order in $M/N$, the $\gamma_{ij}$ of the conformal case can still be used.
This gives that the $\beta_{h_i}=0$ in this approximation.  However, the cancelation of terms
in the $\beta_i$ equations depended crucially on $N_i$ and $N_{j}$ being related.
In the new case the leading order is changed because
 $N_{j}=N_i \left(1+M \alpha_{ij}/{N_i} \right)$ where $\alpha_{ij}$ is a constant which depends
 on the particular quiver and node at hand.
Therefore, the $\beta$ functions associated with the
gauge couplings change in general.  The leading order must still vanish, and this leaves a term
proportional to $M$ left over.
\be
\beta_i=\frac{g_i^3}{16 \pi^2}\frac{M \Sigma \alpha_{ij} T_{r_{ij}}(1-\gamma_{ij})}{1-\frac{g_i^2 N_i}{8\pi^2}}.
\ee

Again, let us turn to our example to be more concrete.  Let $U_1$ and $U_3$ have the same anomalous
dimension $\gamma_1$ and the other fields have the anomalous dimension $\gamma_2$.  The superpotential
beta functions then give that
\be
\left(-3+\left(4+\gamma_{1}+\gamma_{2} \right)\right)=1+\gamma_{1}+\gamma_{2}=0
\ee
(the sum is over 4 terms as there is a quartic superpotential), and the gauge couplings give that
\be
3N-\Sigma N\ft12(1-\gamma_{ij})=0\rightarrow 1+\gamma_{1}+\gamma_{2}=0
\ee
(where the sum is over 4 terms, two arrows in and two out of any given node). 
Note that we have used $T_r(N)= T_r(\bar{N})=1/2$.
In this case, the beta function equations give the same restriction, and we will find this to be
the case in general.  More generically, one can eliminate certain anomalous dimensions
that are related to others by superpotential terms, and then only discuss the remaining anomalous
dimensions.

Perturbing around this fixed point, we find that for nodes 1 and 3 that the new $\beta$ functions
are proportional to
\be
3N-\Sigma (N+M)\ft12(1-\gamma_{ij})=
3N-\Sigma N\ft12(1-\gamma_{ij})-\Sigma M\ft12(1-\gamma_{ij})=-M\left(2-\gamma_1-\gamma_2\right)=-3M
\ee
and that for nodes 2 and 4
\be
3(N+M)-\Sigma (N)\ft12(1-\gamma_{ij})=
3M +3(N)-\Sigma N\ft12(1-\gamma_{ij})+\Sigma M\ft12(1-\gamma_{ij})=3M
\ee
This gives that the new beta functions are non zero, and are proportional to $M$ at leading order.  
%%%%%%%%%%%%%%%%%%%%%%%%%%%%%%%%%%%%%%%%%%%%%%%%%%%%%%%%%%%%%%%%%%%%%%%%%%%%%%%%%%%%%%
\section{Centrally extended finite Heisenberg groups acting on cascading quivers}
\label{meat}
In this section we discuss various examples of cascading theories and explicitly present a group of
discrete transformations acting on these theories. The group in question is a central extension of the
finite Heisenberg group acting in the conformal case.

%%%%%%%%%%%%%%%%%%%%%%%%%%%%%%%%%%%%%%%%%%%%%%%%%%%%%%%%%%%%%%%%%%%%%%%%%%%
\subsection{Orbifolds of $Y^{p,q}$}
%%%%%%%%%%%%%%%%%%%%%%%%%%%%%%%%%%%%%%%%%%%%%%%%%%%%%%%%%%%%%%%%%%%%%%%%%%%

A very interesting class of gauge theories are the quiver gauge theories obtained
as the gauge theory dual of string theory on $AdS_5\times Y^{p,q}$ with 5-form flux.
A very complete discussion of $Y^{p,q}$ spaces is presented in \cite{sasgeom}. The field theory
aspects are presented in \cite{ms,sequiver,friends,toric}.  Inclusion of the fractional branes
and the subsequent cascade on the field theory side was studied in \cite{cells}.  We
start with this set of orbifold models because certain features will be
clearer here than in more symmetric cases.

To be concrete, we work out an example.  Given a conformal quiver gauge theory, one can construct a nonconformal
phase by appropriately changing the rank of some of the gauge groups. The precise recipe involves adding multiples of the
zero eigenvectors of the adjacency matrix as reviewed in section \ref{review}.

First, let us obtain the zero eigenvectors of interest.  
The quivers that we will be dealing with will all respect a ``shift'' symmetry
because there is a fundamental cell that the quiver is built from.  This then
allows us to consider only the adjacency matrix of the sub diagram, identifying
the first and last set of arrows.  For $Y^{6,3}$ we find the sub diagram of $Y^{2,1}$
as
\begin{center}
\includegraphics[width=.25\textwidth]{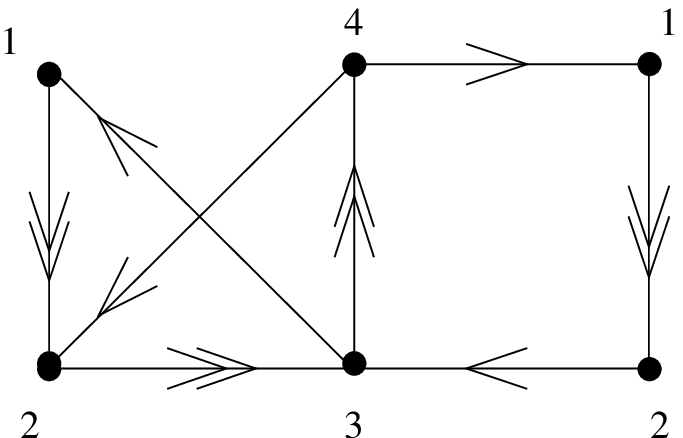}.
\end{center}
with adjacency matrix
\be
\hat{a}=
\begin{pmatrix}
 0 & 2 &-1 &-1 \\
-2 & 0 & 3 &-1 \\
 1 &-3 & 0 & 2 \\
 1 & 1 & -2& 0
\end{pmatrix}.
\ee
The zero eigenvectors of this matrix are
\be
\overrightarrow{v}_1=
\begin{pmatrix}
1 \\
1 \\
1 \\
1
\end{pmatrix}, \quad
\overrightarrow{v}_2=
\begin{pmatrix}
0 \\
2 \\
1 \\
3
\end{pmatrix}.
\ee
To obtain the zero eigenvectors for the $Y^{6,3}$ case, one simply repeats the entries
in the above vectors 3 times.  This process easily generalizes to other quivers
with subcells (except for the cases where $q=0$ or $q=p$, where the unit cells are
the simple $\tau$ and $\sigma$ cells), and so one can see that the $A$ symmetry that
shifts these primitive cells into each other is a symmetry in all cases.

We now turn to the explicit construction of some discrete transformations
$A$, $B$, $C$, and $D$ for the quiver of $Y^{6,3}$ with the rank of the gauge groups shifted
accordingly:
\begin{center}
\includegraphics[width=.5\textwidth]{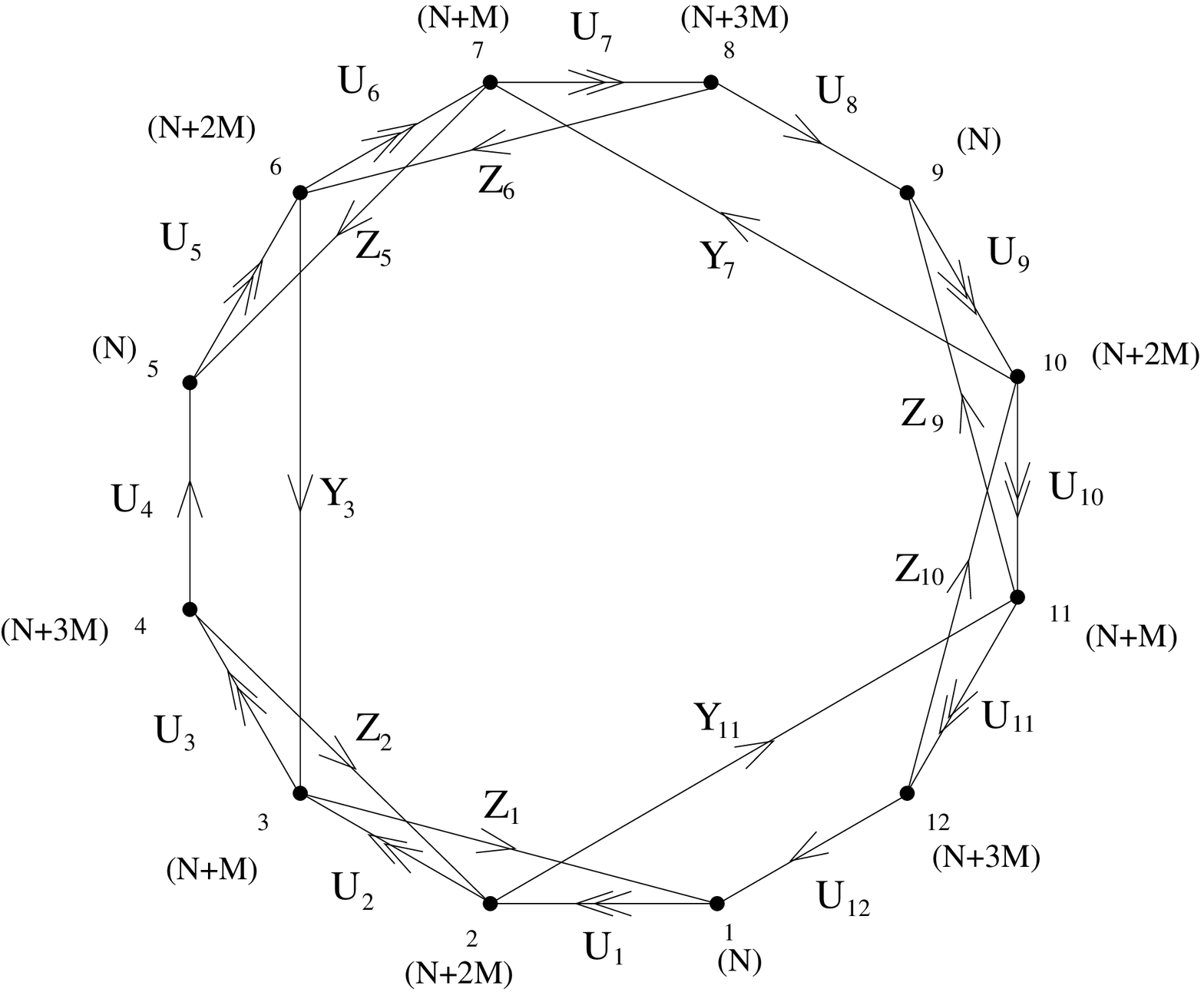}.
\end{center}
The number in parentheses next to the node denotes the rank of that gauge group.
Next, we will recall the results of \cite{heisAlg}, where $M$ is set to $0$.  In
this work, we found a set of transformations $\tilde{A}$,$\tilde{B}$ and $\tilde{C}$:
\be
\tilde{A}:
\begin{array}{ccc}
(1,5,9)&\mapsto& (9,1,5),\\
(2,6,10)&\mapsto& (10,2,6)\\
(3,7,11)&\mapsto& (11,3,7)\\
(4,8,12)&\mapsto& (12,4,8).
\end{array}
\ee
\be
\tilde{B}: U_{i}\mapsto u_i U_{i}
\ee
with
\be
\begin{array}{c c c}
u_1=1 & u_5= \omega^4 & u_9=\omega^8 \\
u_2=\omega & u_6= \omega^3 & u_{10}=\omega^5 \\
u_3=\omega^2 & u_7= \omega^6 & u_{11}=\omega^{10} \\
u_4=\omega^{-1} & u_8= \omega^{-3} & u_{12}=\omega^{-35} \\
\end{array},
\ee
and
\be
\tilde{C}: U_{i}\mapsto u_i U_{i}
\ee
with
\be
\begin{array}{c c c}
u_1=\omega^4 & u_5= \omega^4 & u_9=\omega^4 \\
u_2=\omega^2 & u_6= \omega^2 & u_{10}=\omega^2 \\
u_3=\omega^4 & u_7= \omega^4 & u_{11}=\omega^{4} \\
u_4=\omega^{-2} & u_8= \omega^{-2} & u_{12}=\omega^{-26} \\
\end{array}.
\ee
and where $\omega^{3N}=1$.
These satisfy a finite Heisenberg group structure
\be
\tilde{A}\tilde{B}=\tilde{B}\tilde{A}\tilde{C},
\quad \tilde{A}\tilde{C}=\tilde{C}\tilde{A},
\quad \tilde{A}^3=\tilde{B}^3=\tilde{C}^3=1
\ee
and $C$ commutes with all generators above.  These equations are to
be read up to members of the center of the gauge group.
The above transformations also satisfy the anomaly
cancelation conditions
\bea
\left(\frac{u_{12} u_1}{u_2}\right)^N &=&1, \quad
\left(\frac{u_1 u_2}{u_{11} u_{12} u_{3}}\right)^N =1, \nn \\
\left(\frac{u_2 u_3}{u_1 u_4 u_5}\right)^N &=&1, \quad
\left(\frac{u_3 u_4}{u_2}\right)^N =1, \nn\\
\left(\frac{u_4 u_5}{u_6}\right)^N &=&1, \quad
\left(\frac{u_5 u_6}{u_3 u_4 u_7}\right)^N =1, \label{anomCondY63Conf} \\
\left(\frac{u_6 u_7}{u_5 u_8 u_9}\right)^N &=&1, \quad
\left(\frac{u_7 u_8}{u_6}\right)^N =1, \nn\\
\left(\frac{u_8 u_9}{u_{10}}\right)^N &=&1, \quad
\left(\frac{u_9 u_{10}}{u_7 u_8 u_{11}}\right)^N =1, \nn\\
\left(\frac{u_{10} u_{11}}{u_{9}u_{12}u_{1}}\right)^N &=&1,\quad
\left(\frac{u_{11} u_{12}}{u_{10}}\right)^N =1\nn.
\eea
We will now generalize these to the non-conformal case.
First, we may note that because the $A$ symmetry shifts gauge groups
of the same rank into each other and so this remains a symmetry
in the non conformal case.  Next, we recall that
the $B$ and $C$ symmetries were constructed just so that they
were $3^{\rm rd}$ roots of members of the center of the gauge
group.  The particular members were labeled by assigning each gauge
group a number
\begin{center}
\includegraphics[width=\textwidth]{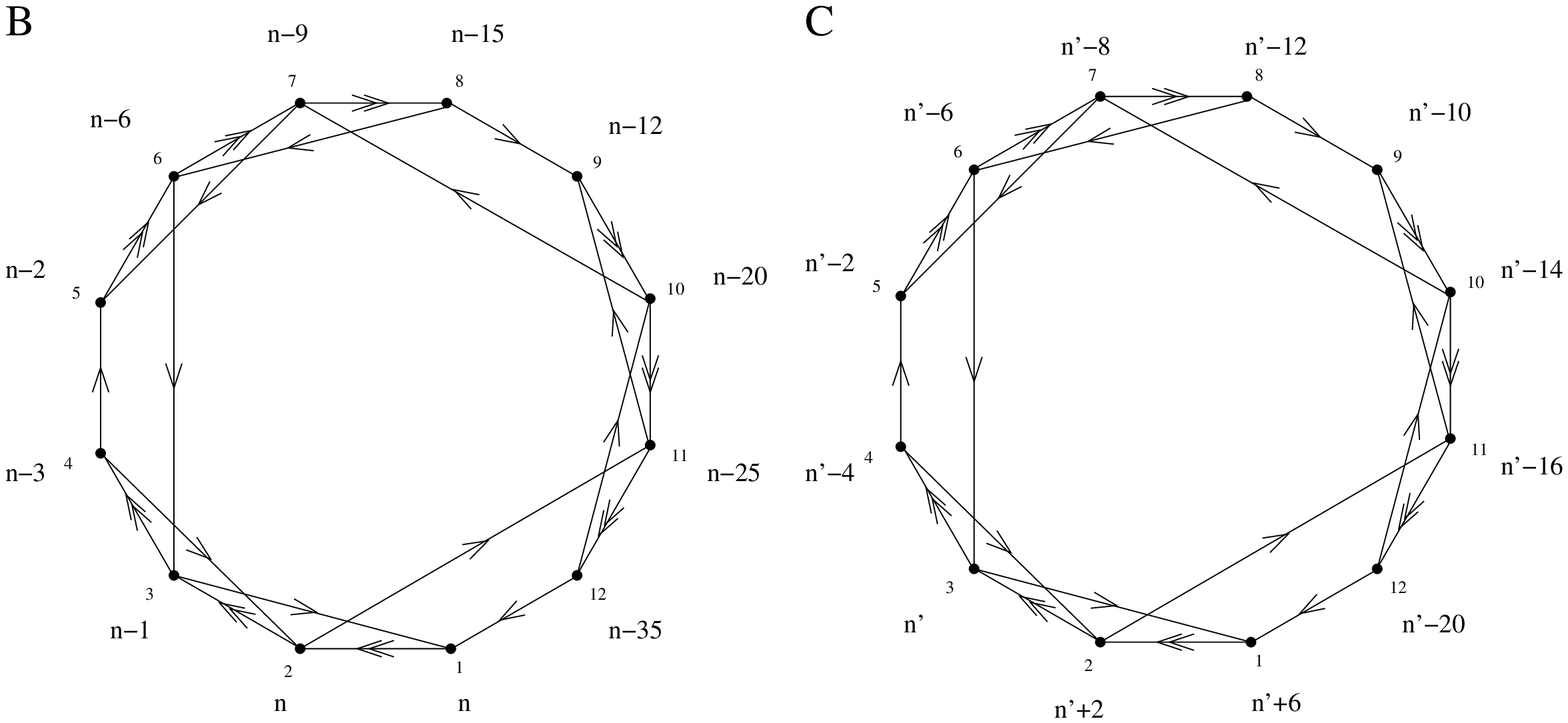}
\end{center}
and then fields were rephased as $U_i\rightarrow \omega^{(n_{i})}\omega^{(-n_{i+1})}U_i$.
This then gives that these operators to the $3^{\rm rd}$ power are automatically in
the center of the gauge group (as $\omega^3$ is an $N^{\rm th}$ root of unity).
Note that $n$ and $n'$ are arbitrary in the conformal case above because each gauge group
has the same $\omega$ associated with it, as the gauge groups are the same rank.

We will now set about generalizing this to the non-conformal case.  For $M\neq0$ we define
the useful quantity
\be
\lambda\equiv \frac{N}{M}.
\ee
We now associate different $\omega$'s to each gauge group.  We
want that they will eventually be related to the center of the
gauge group, and so we define
\bea
\omega_0 &=& \omega^{\frac{\lambda+3}{\lambda}} \nn \\
\omega_1 &=& \omega^{\frac{\lambda+3}{\lambda+1}} \nn \\
\omega_2 &=& \omega^{\frac{\lambda+3}{\lambda+2}}  \\
\omega_3 &=& \omega^{\frac{\lambda+3}{\lambda+3}}=\omega, \nn
\eea
i.e. we have solved the $(3(\lambda+i)M)^{\rm th}$ roots of unity for $i=0,1,2$
in terms of the  $(3(\lambda+3)M)^{\rm th}$ root of unity, which we call $\omega$.
Therefore, instead of associating the same $\omega$ with each gauge group,
we associate the phase $\omega_i$ with a gauge group with rank $(\lambda+i)M$.
Now that different phases are assigned to different gauge groups, the overall
numbers $n$ and $n'$ do affect the final result.  In our example we will take
$n=0$ and then adjust $n'$ so that $u^{B}_5=u^{C}_5$.  This will give that the $C$
operation properly ``undoes'' the affect of shift symmetry $A$.  The solution to
this restriction is $n'=0$.

Using the above construction, we find
\be
A=\tilde{A}:
\begin{array}{ccc}
(1,5,9)&\mapsto& (9,1,5),\\
(2,6,10)&\mapsto& (10,2,6)\\
(3,7,11)&\mapsto& (11,3,7)\\
(4,8,12)&\mapsto& (12,4,8),
\end{array}
\ee
and
\be
B: U_{i}\mapsto u_i U_{i},\;
\ee
with
\be
\begin{array}{c c c}
u_1=1 &
u_5= \omega^{4\frac{(\lambda+3)(\lambda-1)}{(\lambda)(\lambda+2)}} &
u_9=\omega^{8\frac{(\lambda+3)(\lambda-1)}{(\lambda)(\lambda+2)}} \\
u_2=\omega^{\frac{\lambda+3}{\lambda+1}} &
u_6= \omega^{3\frac{(\lambda+3)(\lambda+4)}{(\lambda+2)(\lambda+1)}}  &
u_{10}=\omega^{5\frac{(\lambda+3)(\lambda+6)}{(\lambda+2)(\lambda+1)}} \\
u_3=\omega^{2\frac{\lambda}{\lambda+1}} &
u_7= \omega^{6\frac{\lambda-2}{\lambda+1}} &
u_{11}=\omega^{10\frac{\lambda-4}{\lambda+1}}  \\
u_4=\omega^{-\frac{(\lambda-6)}{\lambda}} &
u_8= \omega^{-3\frac{(\lambda-12)}{\lambda}} &
u_{12}=\omega^{-35} \\
\end{array},
\ee
and
\be
C: U_{i}\mapsto u_i U_{i}
\ee
with
\be
\begin{array}{c c c}
u_1=\omega^{4\frac{(\lambda+3)(\lambda+3)}{(\lambda)(\lambda+2)}} &
u_5=\omega^{4\frac{(\lambda+3)(\lambda-1)}{(\lambda)(\lambda+2)}}  &
u_9=\omega^{4\frac{(\lambda+3)(\lambda-5)}{(\lambda)(\lambda+2)}}  \\
u_2=\omega^{2\frac{(\lambda+3)(\lambda+1)}{(\lambda+2)(\lambda+1)}}&
u_6=\omega^{2\frac{(\lambda+3)(\lambda+5)}{(\lambda+2)(\lambda+1)}} &
u_{10}=\omega^{2\frac{(\lambda+3)(\lambda+9)}{(\lambda+2)(\lambda+1)}} \\
u_3=\omega^{4\frac{(\lambda+1)}{(\lambda+1)}}  &
u_7=\omega^{4\frac{(\lambda-3)}{(\lambda+1)}} &
u_{11}=\omega^{4\frac{(\lambda-7)}{(\lambda+1)}} \\
u_4=\omega^{-2\frac{(\lambda-3)}{\lambda}} &
u_8= \omega^{-2\frac{(\lambda-15)}{\lambda}} &
u_{12}=\omega^{-2\frac{(13\lambda+9)}{\lambda}} \\
\end{array}.
\ee
We find that the above transformations indeed satisfy the
new anomaly cancelation conditions
\bea
\left(\frac{u_{12}^{\lambda+3} u_1^{\lambda+3}}{u_2^{\lambda+1}}\right)^M &=&1, \quad
\left(\frac{u_1^{\lambda-1} u_2^{\lambda-1}}{u_{11}^{\lambda+1} u_{12}^{\lambda+1} u_{3}^{\lambda+3}}\right)^M =1, \nn \\
\left(\frac{u_2^{\lambda+4} u_3^{\lambda+4}}{u_1^{\lambda} u_4^{\lambda+2} u_5^{\lambda+2}}\right)^M &=&1, \quad
\left(\frac{u_3^{\lambda} u_4^{\lambda}}{u_2^{\lambda+2}}\right)^M =1, \nn\\
\left(\frac{u_4^{\lambda+3} u_5^{\lambda+3}}{u_6^{\lambda+1}}\right)^M &=&1, \quad
\left(\frac{u_5^{\lambda-1} u_6^{\lambda-1}}{u_3^{\lambda+1} u_4^{\lambda+1} u_7^{\lambda+3}}\right)^M =1, \label{anomCond} \\
\left(\frac{u_6^{\lambda+4} u_7^{\lambda+4}}{u_5^{\lambda} u_8^{\lambda+2} u_9^{\lambda+2}}\right)^M &=&1, \quad
\left(\frac{u_7^{\lambda} u_8^{\lambda}}{u_6^{\lambda+2}}\right)^M =1, \nn\\
\left(\frac{u_8^{\lambda+3} u_9^{\lambda+3}}{u_{10}^{\lambda+1}}\right)^M &=&1, \quad
\left(\frac{u_9^{\lambda-1} u_{10}^{\lambda-1}}{u_7^{\lambda+1} u_8^{\lambda+1} u_{11}^{\lambda+3}}\right)^M =1, \nn\\
\left(\frac{u_{10}^{\lambda+4} u_{11}^{\lambda+4}}{u_{9}^{\lambda}u_{12}^{\lambda+2}u_{1}^{\lambda+2}}\right)^M &=&1,\quad
\left(\frac{u_{11}^{\lambda} u_{12}^{\lambda}}{u_{10}^{\lambda+2}}\right)^M =1\nn.
\eea
This works in much the same way as the conformal case, where the first and last line
impose the condition $\omega^{3(\lambda+3)M}=1$, and
the other equalities are satisfied trivially before raised to the $M^{\rm th}$ power
(actually, a $6$ could appear instead
of 3 as the overall factor in the exponent, however we wish to find operations that
when raised to the $3^{\rm rd}$ power give $1$, or a member of the center of the
gauge group).

The above relations satisfy
\be
AB=BAC\times M \equiv BAC, \quad A^3=B^3=C^3=1
\ee
where $M$ is in the center of the gauge group, given by the element
\begin{center}
\includegraphics[width=.5\textwidth]{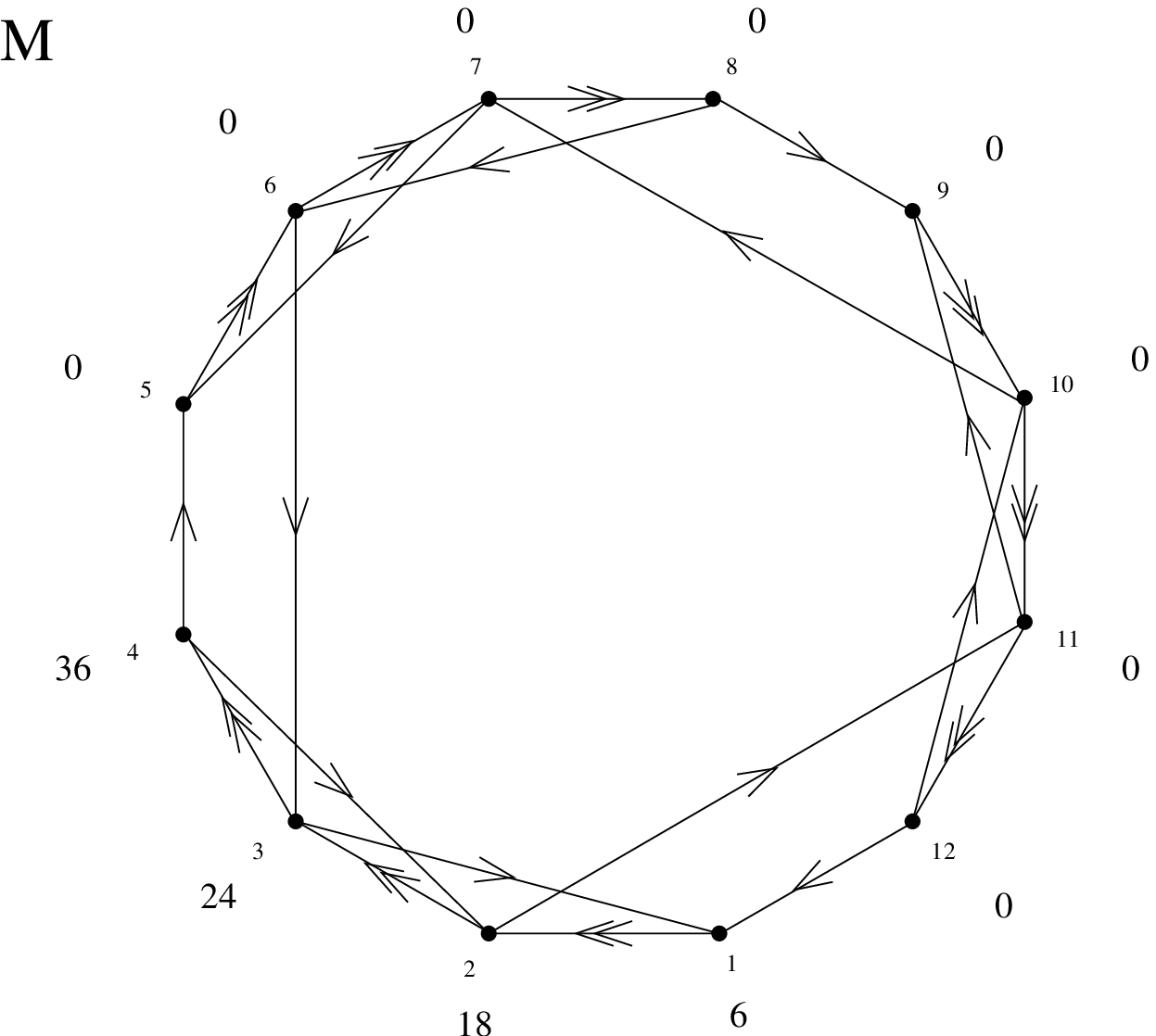}.
\end{center}
This is in the center because all of the numbers appearing above are
divisible by $3$.

We now turn to the commutation of $C$ with other generators.  It is trivial
that $C$ commutes with $B$ because they are both diagonal.  However, we now
find that
\be
AC=CAD
\ee
with $D$ defined as
\be
D: U_{i}\mapsto u_i U_{i}
\ee
with
\be
\begin{array}{c c c}
u_1=\alpha^8&
u_5=\alpha^{-4}  &
u_9=\alpha^{-4}  \\
u_2=\beta^{-8}&
u_6=\beta^{4}  &
u_{10}=\beta^{4}  \\
u_2=\gamma^{8}&
u_6=\gamma^{-4}  &
u_{10}=\gamma^{-4}  \\
u_4=\omega^{24\frac{(\lambda+1)}{\lambda}} &
u_8= \omega^{24\frac{1}{\lambda}} &
u_{12}=\omega^{-24\frac{(\lambda+2)}{\lambda}} \\
\end{array}.
\ee
where we have defined the useful quantities
\be
\alpha=\omega^{4\frac{(\lambda+3)}{(\lambda)(\lambda+2)}},\quad
\beta=\omega^{2\frac{(\lambda+3)}{(\lambda+2)(\lambda+1)}}, \quad
\gamma=\omega^{4\frac{1}{(\lambda+1)}}.
\ee
The powers appearing in the definition of $\alpha, \beta,$ and $\gamma$ are
related to the rank of the gauge groups that the $(1,5,9),(2,6,9)$ and $(3,7,10)$
fields run between.  Note that because the operator $C$ satisfies the
anomaly conditions, and $A$ simply permutes them, then $D$ automatically
satisfies the anomaly conditions.  Also, $D$ commutes trivially with all generators
except $A$, and so we check that
\be
AD=DA\times M'
\ee
and we find that $M'$ is in the center of the gauge group
corresponding to the element
\begin{center}
\includegraphics[width=.5\textwidth]{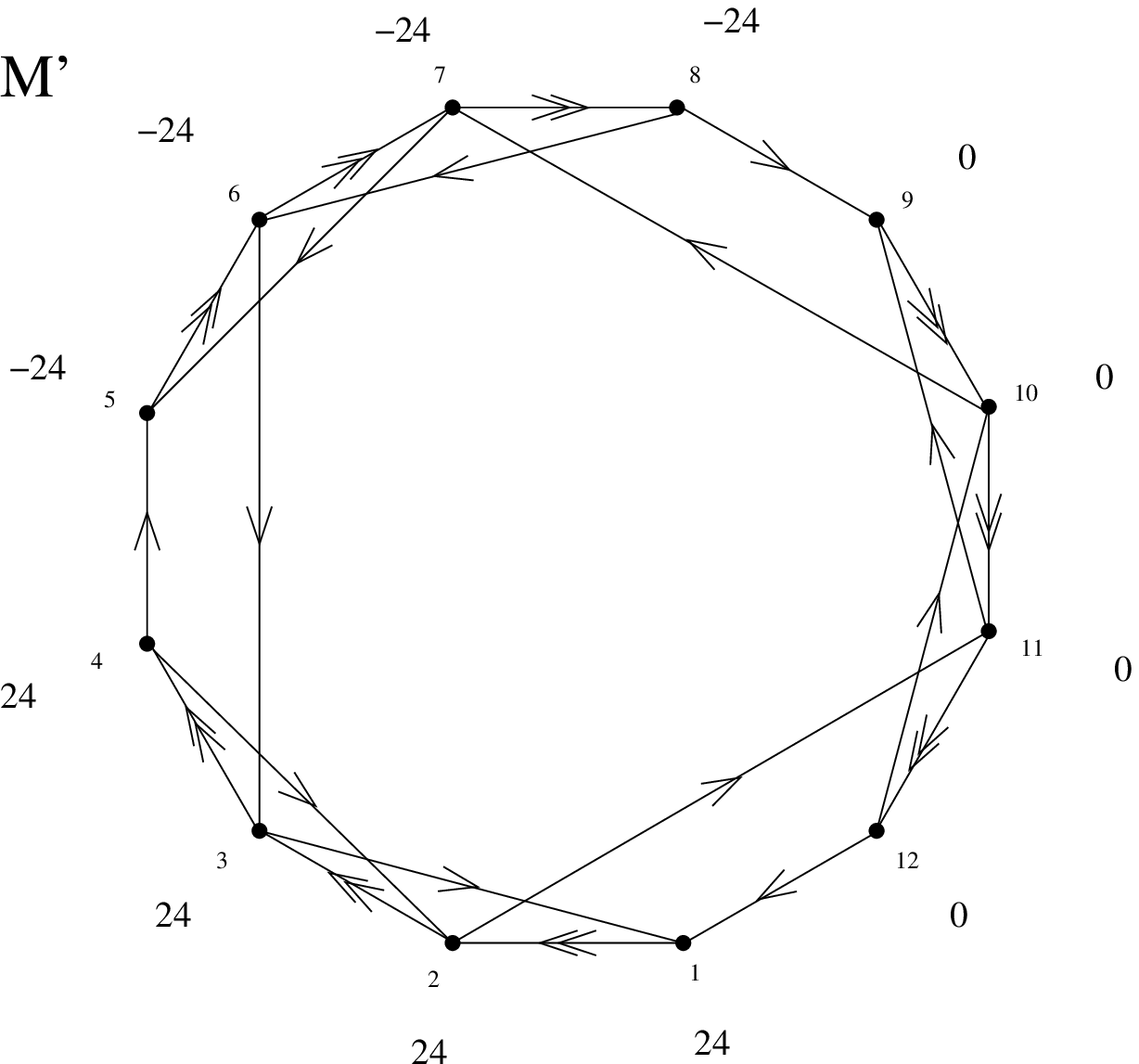}.
\end{center}
We now comment on $D$.  First, one can find that $D$ is generated
by
\begin{center}
\includegraphics[width=.5\textwidth]{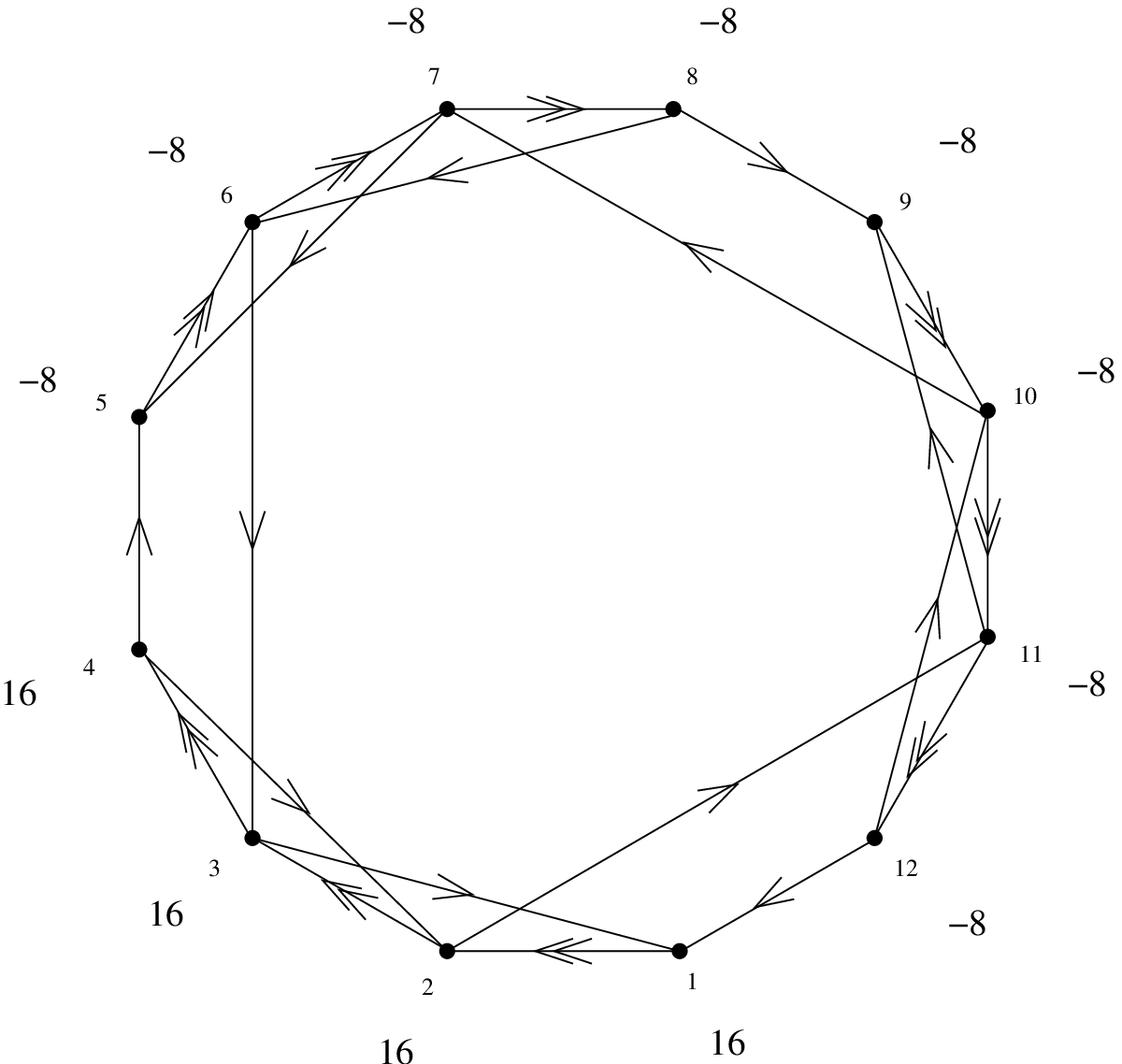}.
\end{center}
The crucial difference between the conformal and non conformal cases
is now clear: in the conformal case we could raise all of the
integer powers by $8$ appearing above.  Then, only $0$ and $24$
would appear at each node, both of which are divisible by $3$.
Another way to say this is that all the numbers appearing above
are congruent to $1$ mod $3$.  Thus, $D$ is gauge equivilant
to the case where one simply enters $1$ for all nodes.  This
is a trivial operator in the conformal case because the $\omega$
associated with each gauge group is the same, however in the nonconformal case the
$\omega_i$ associated with each node is different.

Also of interest is the fact that $D$ is simply a ${\zet}_3$ subgroup
of an entire $U(1)$.  This is because all of the anomaly cancelation
conditions are met trivially (before raising to the $M^{\rm th}$ power).
Hence, there is no ``wrapping'' condition that requires the phases
appearing in $D$ to be any particular root of unity, and so are
arbitrary $U(1)$ phases.  Curiously, there is another symmetry that
is also a ${\zet}_3$ subgroup of a $U(1)$:
\be
E: U_{i}\mapsto u_i U_{i}
\ee
with
\be
\begin{array}{c c c}
u_1=\omega^{\frac{-2(\lambda+3)}{(\lambda+2)}}&
u_5=\omega^{\frac{-2(\lambda+3)}{(\lambda+2)}}  &
u_9=\omega^{\frac{-2(\lambda+3)}{(\lambda+2)}} \\
u_2=\omega^{\frac{\lambda(\lambda+3)}{(\lambda+1)(\lambda+2)}}&
u_6=\omega^{\frac{\lambda(\lambda+3)}{(\lambda+1)(\lambda+2)}}  &
u_{10}=\omega^{\frac{\lambda(\lambda+3)}{(\lambda+1)(\lambda+2)}}  \\
u_2=\omega^{\frac{-2\lambda}{(\lambda+1)}}&
u_6=\omega^{\frac{-2\lambda}{(\lambda+1)}}  &
u_{10}=\omega^{\frac{-2\lambda}{(\lambda+1)}}  \\
u_4=\omega^{3} &
u_8= \omega^{3} &
u_{12}=\omega^{3} \\
\end{array}.
\ee
This can be generated from the diagram
\begin{center}
\includegraphics[width=.5\textwidth]{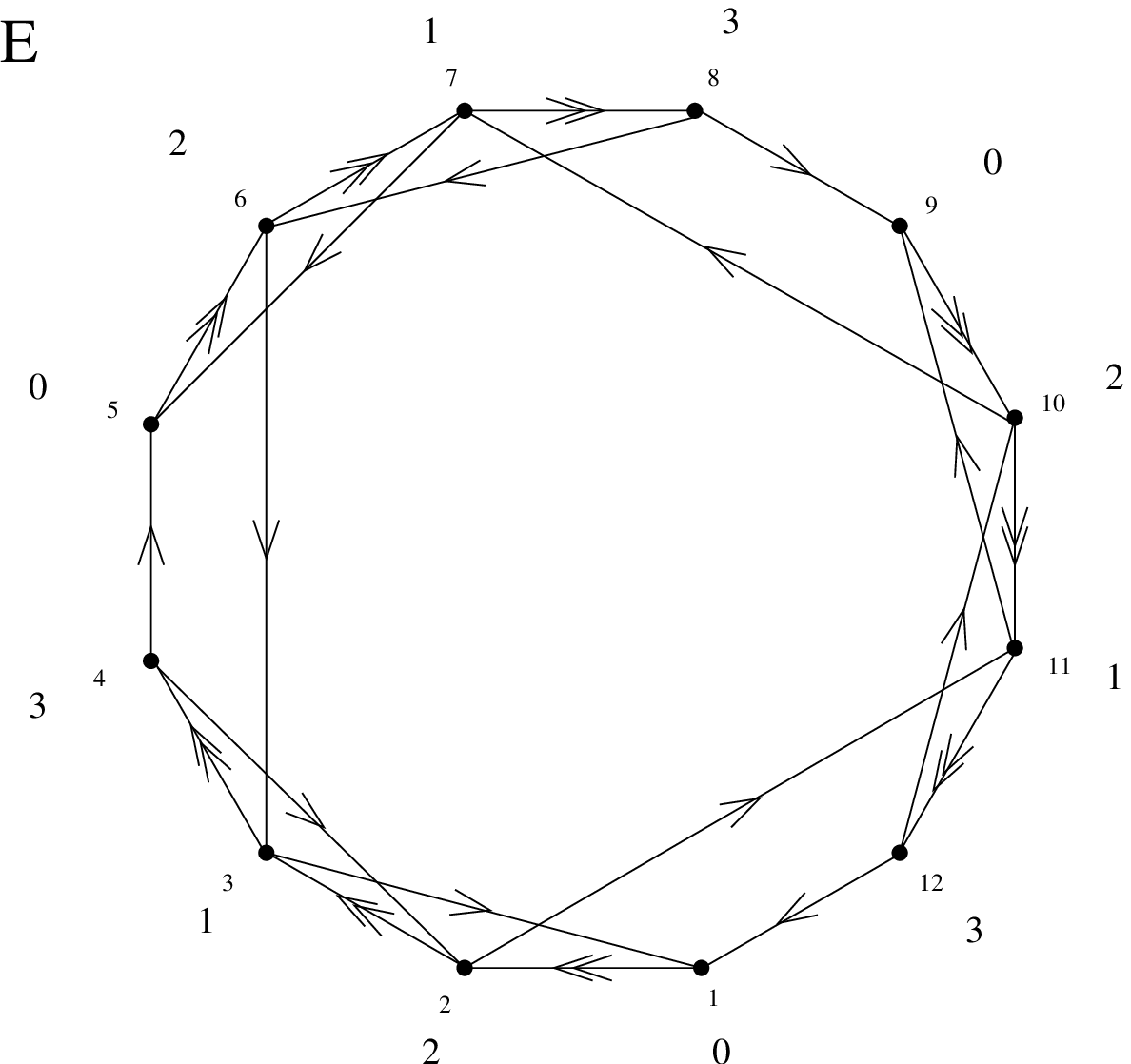}.
\end{center}
One should note that the assignments of $D$ and $E$ correspond
exactly to the vectors that determine the rank of the gauge
groups: D for the coefficient of $N$ and $E$ for the coefficient
of $M$ in the gauge groups.  One should also notice that while $D$ is trivial
in the conformal case, $E$ is not.  The conformal
case has an additional symmetry to those found in \cite{heisAlg},
making the group generators satisfy that of
$\tilde{H} \times\zet_3$ using only $E$, or $\tilde{H}\times U(1)$ using
the $U(1)$ that $E$ is a subgroup of (denoting the Heisenberg
group as $\tilde{H}$).
In summary, the algebra that we find is
\bea
&& AB=BAC,\quad AC=CAD, \quad AD=DA, \quad \{B,C,D,E \} \; {\rm commute} \nn \\
&& A^3=B^3=C^3=D^3=E^3=1.
\eea
We will refer to the finite Heisenberg group as ${\rm Heis}(\mathbb{Z}_3\times \mathbb{Z}_3)$ and the above
group with $E$ removed as $\bar{H}_3$ (the $3$ to denotes that we are really talking about groups modulo
$3$, and we remove $E$ because it always appears as an uninteresting direct product factor).
We note that an arbitrary element of ${H}_3$ can be written
$A^a B^b C^c D^d$ with $a,b,c,d \in \{0,1,2\}$.  Likewise, an arbitrary element of ${\rm Heis}(\mathbb{Z}_3\times \mathbb{Z}_3)$
may be
written as $\tilde{A}^a \tilde{B}^b \tilde{C}^c$ again with $a,b,c\in \{0,1,2\}$.
One may therefore view ${H}_3$ as a central extension
of the ${\rm Heis}(\mathbb{Z}_3\times \mathbb{Z}_3)$.  To be explicit, we take
\[\bfig
\morphism(0,200)|a|/{->}/<500,0>[\Id`{\zet_3(D)};f_1]
\morphism(500,200)|a|/{->}/<500,0>[{\zet_3(D)}`{H}_3;f_2]
\morphism(1000,200)|a|/{->}/<500,0>[{H}_3`{\rm Heis}(\mathbb{Z}_3\times \mathbb{Z}_3);f_3]
\morphism(1500,200)|a|/{->}/<500,0>[{\rm Heis}(\mathbb{Z}_3\times \mathbb{Z}_3)`\Id;f_4]
\efig\]
where the $\zet_3$ is the group generated by $D$.  Therefore, we take the maps

\bea
f_1(\Id)&=&D^0 \nn \\
f_2(D^d)&=&D^d \nn \\
f_3(A^a B^b C^c D^d) &=& \tilde{A}^a \tilde{B}^b \tilde{C}^c \\
f_4(\tilde{A}^a \tilde{B}^b \tilde{C}^c) &= &\Id \nn
\eea
and find that
\bea
{\rm Ker}(f_1)&=& \Id \nn \\
{\rm Im}(f_1) &=& D^0 \nn \\
{\rm Ker}(f_2)&=& D^0 \nn \\
{\rm Im}(f_2) &=& D^d  \\
{\rm Ker}(f_3)&=& D^d \nn \\
{\rm Im}(f_3) &=& \tilde{A}^a \tilde{B}^b \tilde{C}^c \nn \\
{\rm Ker}(f_4)&=& \tilde{A}^a \tilde{B}^b \tilde{C}^c \nn \\
{\rm Im}(f_4) &=& \Id.
\eea
So, the above is an exact sequence of homomorphisms, and
further $D^d$ is in the center of ${H}$.

One may worry about the ``internal'' fields in the diagram.  These, however
will satisfy the algebra above in the same way.  Let us take as an
example the $Z_a$ fields appearing in quiver.  The scalings $z_a$ are always
read from the superpotential constraints, and so the $B$ $C$ and $D$ operators
are determined directly.  Let us take an example.  We may show that because
$AB=BAC$ is satisfied for the $U$ fields that it is also satisfied for the
$Z$ fields.  We will refer to the scalings associated with the $B$ operation
acting on the $U$ fields as $u_a^B$, and likewise for other fields and operations.
We find
\bea
(AB)Z_a &=& (u^B_{a+4} u^B_{a+5})^{-1} Z_{a+4} \nn \\
(BAC) Z_a &=& (u^B_{a} u^B_{a+1}u^C_{a+4} u^C_{a+5})^{-1} Z_{a+4}
\eea
Next, because we have already solved the $U$ problem, we have that
\be
u^B_{a}u^C_{a+4}\times (G_{a+4,a+5})=u^B_{a+4} u^B_{a+5} \label{Usect}
\ee
where $G_{a+4,a+5}$ denotes the component of the center of the gauge group
that rephases nodes $a+4$ and $a+5$, i.e. those that affect $U_{a+4}$.  From this,
we determine that
\bea
(AB)Z_a &=& BAC (G_{a,a+1}G_{a+1,a+2})^{-1}Z_a \label{Zsect}
\eea
In the above manipulations one must shift $a$ down by $4$ in (\ref{Zsect})
relative to the $a$ that appears in (\ref{Usect})
so that both sides match after $A$ shifts $a$ up by $4$ on the RHS of (\ref{Zsect}).
We note that $G_{a,b}=\omega_{i}^{n_a}\omega_j^{-n_{b}}$ where we use
$i$ to denote the rank of the gauge group as $(\lambda +i)M$ at node $a$ and $j$
likewise denotes the rank of the gauge group at node $b$.  Thus,
$G_{a,b}G_{b,c}=G_{a,c}$.  Therefore, we find that
\bea
(AB)Z_a &=& BAC (G_{a,a+2})^{-1}Z_a.
\eea
as expected.  The inverse power shows up precisely because Z is oriented from node $a+2$
to node $a$ rather than the ``forward'' direction.  Thus, in the $Z$ sector
$AB=BAC$ follows from the relation $AB=BAC$ in the $U$ sector.  This is likewise true for
the fields $Y$ that enter in quartic superpotential terms, only now $3$ of the
$G$ scalings ``telescope'' to become the single one needed.  The fact that
$B^3=1$ in the $Z$ sector is also obvious.  All equations
are of this form and so the $Z$ and $Y$ sectors follow automatically.  For
this reason we will not treat the internal lines in the remainder of the
paper, knowing that the commutation relation follow in a trivial manner
given the commutations of the fields on the ``outside'' of the quiver.

%%%%%%%%%%%%%%%%%%%%%%%%%%%%%%%%%%%%%%%%%%%%%%%%%%%%%%%%%%%%%%%%%%%%%%%%%%
\subsection{Orbifolds of $S^{5}=Y^{p,p}$}
%%%%%%%%%%%%%%%%%%%%%%%%%%%%%%%%%%%%%%%%%%%%%%%%%%%%%%%%%%%%%%%%%%%%%%%%%%%%

A clear set of orbifold examples to explore are those of $S^{5}$.  We will
concentrate particularly on orbifolds that correspond to $Y^{p,p}$.  These
theories have $2p$ gauge groups, and so are ${\zet}_{2p}$ orbifolds of the
sphere.  We find it easier to work an example, and then display the generic
features.  We will concentrate on the $AdS_5 \times Y^{6,6}$ geometry with
imaginary self dual three form turned on.  The field theory dual to this
string background is given by the quiver
\begin{center}
\includegraphics[width=.5\textwidth]{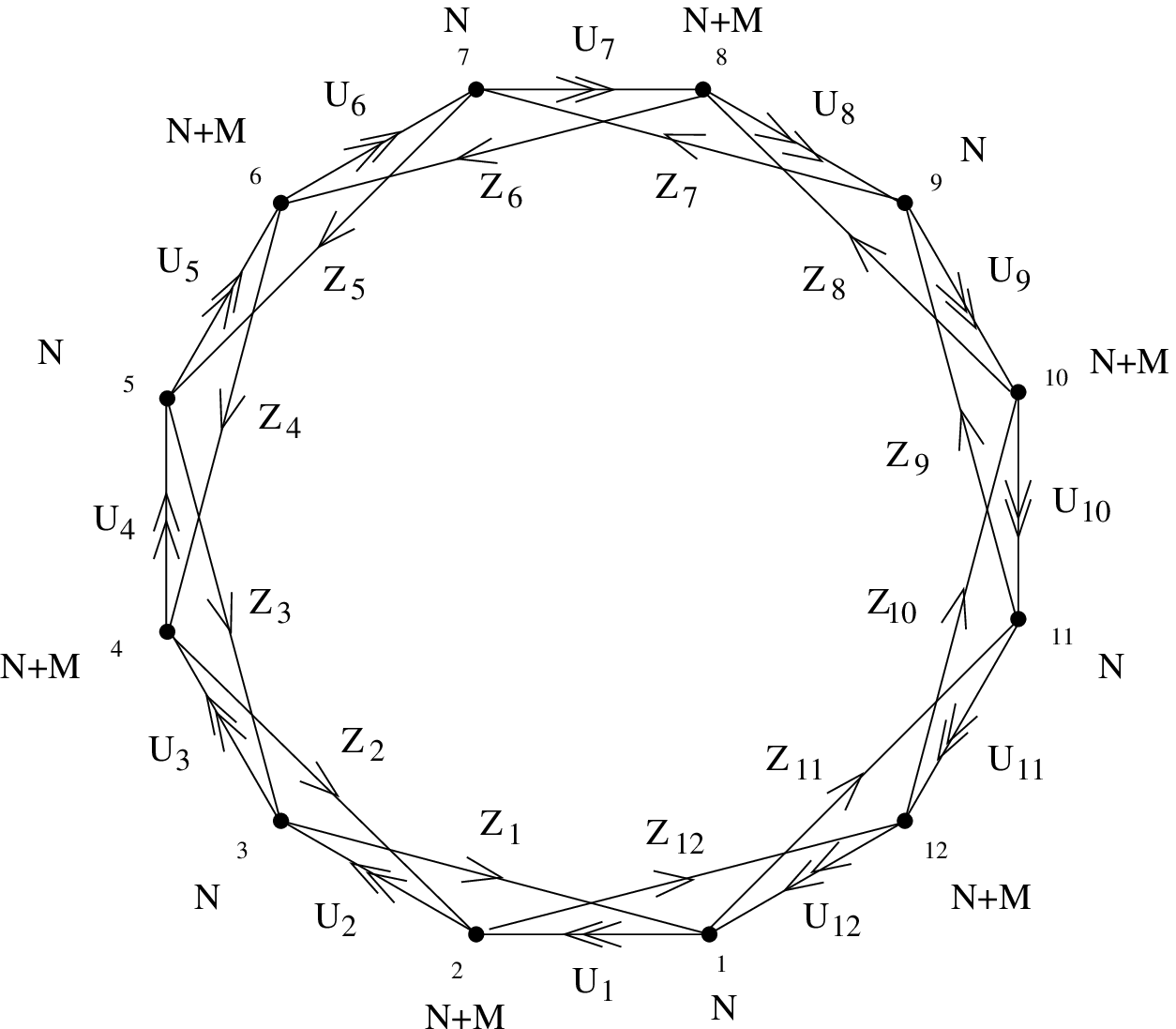}.
\end{center}
We find that the above field theory has the following shift symmetry
\be
A:
\begin{array}{ccc}
(1,3,5,7,9,11) & \mapsto &(3,5,7,9,11,1), \\
(2,4,6,8,10,12)& \mapsto &(4,6,8,10,12,2)\\
\end{array}
\ee
and the following rephasing symmetries
\be
B: U_{i}\mapsto u_i U_{i}
\ee
with
\be
\begin{array}{c c c c c c}
u_1=1 & u_3=\omega & u_5=\omega^{\frac{\lambda-2}{\lambda}} & u_7=\omega^{\frac{3\lambda-6}{\lambda}}
& u_9=\omega^{\frac{4\lambda-12}{\lambda}} & u_{11}=\omega^{\frac{5\lambda-20}{\lambda}} \\
u_2=1 & u_4=\omega^{\frac{\lambda+2}{\lambda}} & u_6=\omega^{\frac{2\lambda+6}{\lambda}} &
u_8=\omega^{\frac{3\lambda+12}{\lambda}} & u_{10}=\omega^{\frac{4\lambda+20}{\lambda}} &
u_{12}=\omega^{-\frac{25\lambda}{\lambda}},
\end{array}
\ee
and
\be
C: U_{i}\mapsto u_i U_{i}
\ee
with
\be
\begin{array}{c c c c c c}
u_1=\omega^{\frac{\lambda+2}{\lambda}} & u_3=\omega & u_5=\omega^{\frac{\lambda-2}{\lambda}} &
u_7=\omega^{\frac{\lambda-4}{\lambda}}
& u_9=\omega^{\frac{\lambda-6}{\lambda}} & u_{11}=\omega^{\frac{\lambda-8}{\lambda}} \\
u_2=\omega & u_4=\omega^{\frac{\lambda+2}{\lambda}} &
u_6=\omega^{\frac{\lambda+4}{\lambda}} &
u_8=\omega^{\frac{\lambda+6}{\lambda}} & u_{10}=\omega^{\frac{\lambda+8}{\lambda}} &
u_{12}=\omega^{-\frac{11\lambda+2}{\lambda}},
\end{array}
\ee
and
\be
D: U_{i}\mapsto u_i U_{i}
\ee
with
\be
\begin{array}{c c c c c c}
u_1=\omega^{\frac{1}{\lambda}} & u_3=\omega^{\frac{1}{\lambda}} & u_5=\omega^{\frac{1}{\lambda}} &
u_7=\omega^{\frac{1}{\lambda}}
& u_9=\omega^{\frac{1}{\lambda}} & u_{11}=\omega^{\frac{1}{\lambda}} \\
u_2=\omega^{-\frac{1}{\lambda}} & u_4=\omega^{-\frac{1}{\lambda}}  &
u_6=\omega^{-\frac{1}{\lambda}}  &
u_8=\omega^{-\frac{1}{\lambda}} & u_{10}=\omega^{-\frac{1}{\lambda}}  &
u_{12}=\omega^{-\frac{1}{\lambda}} ,
\end{array}
\ee
and
\be
E: U_{i}\mapsto u_i U_{i}
\ee
with
\be
\begin{array}{c c c c c c}
u_1=\omega^{-1} & u_3=\omega^{-1} & u_5=\omega^{-1} &
u_7=\omega^{-1}
& u_9=\omega^{-1} & u_{11}=\omega^{-1} \\
u_2=\omega & u_4=\omega  &
u_6=\omega  &
u_8=\omega & u_{10}=\omega  &
u_{12}=\omega ,
\end{array}
\ee
and where $\omega^{6(\lambda+1)M}=1$.

One should note that while we have defined $C$ the operator $C^{\fft12}$ is also
a well defined symmetry of the system (but not further roots).  Likewise, $D$ and
$E$ are actually the ${\zet}_3$ subgroup of two full $U(1)$ symmetries.  These
symmetries obey the following properties
\bea
&& AB=BAC,\quad AC=CAD^{-2}, \quad AD=DA, \quad AE=EA, \quad \{B,C,D,E\}\; {\rm commute},\nn  \\
&& A^6=B^6=C^6=D^6=E^6=1
\eea
where the equal signs are read up to the center of the gauge group.  One may
read these transformations as coming from the diagrams
\begin{center}
\includegraphics[width=\textwidth]{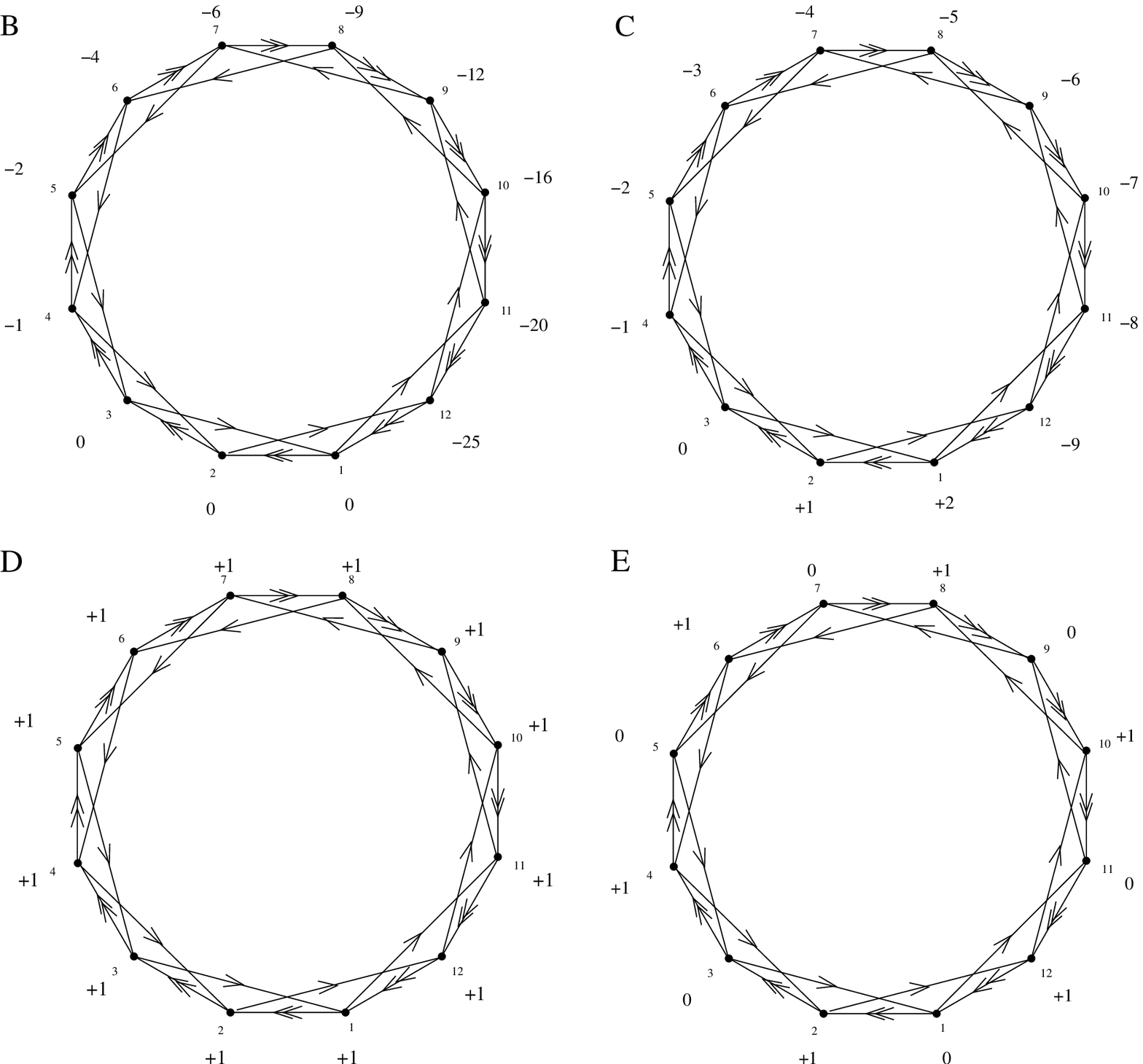}
\end{center}
and one should note that again $D$ and $E$ are related to the
zero eigenvectors of the adjacency matrix.  Finally, we note here that while the
conformal case had operators that were order 12, here they are order
6.  This change comes about because the new gauge ranks do not respect
the original $A$ symmetry.  The only candidate $A$ operation would be
a map between a quiver where odd nodes have rank $N+M$ and even nodes
have rank $N$ to a quiver where odd nodes have rank $N$ and even
nodes have rank $N+M$.  These, however, are not related by Seiberg
duality, as this will shift the ranks of the groups by total factors
of $2M$.

%%%%%%%%%%%%%%%%%%%%%%%%%%%%%%%%%%%%%%%%%%%%%%%%%%%%%%%%%%%%%%%%%%%%%%%%%%%
\section{String theory interpretation}\label{string}
%%%%%%%%%%%%%%%%%%%%%%%%%%%%%%%%%%%%%%%%%%%%%%%%%%%%%%%%%%%%%%%%%%%%%%%%%%%
Various aspects of the physics on this nonconformal quiver gauge theories have been 
approached from the AdS/CFT point of view, starting with the original idea of adding three-form 
flux in \cite{kn} and including the interpretation of the supergravity background as describing 
a cascade of Seiberg dualities
in  \cite{ks}. In the concrete case of $Y^{p,q}$ spaces some results
were presented in \cite{cells}. A further generalization was discussed
in \cite{ypqdef}.  Most of the discussion of the field theory in
\cite{cells} is based on identifications from \cite{chaos}, where
the general role of the wrapped D5 branes was elucidated.

The addition of the possible D5 branes is constrained by chiral
anomaly cancelation. As mentioned in section 2, adding D5 branes
is correlated with the existence of null eigenvectors of the adjacency
matrix.  The wrapped D5 branes are described as $G_3$ flux in the
supergravity background.

To keep some control of the result near the conformal fixed point
we will assume that $M\ll N$. In the supergravity side this limit
allows to neglect the backreaction of the the three-form flux. When
taken into consideration this backreaction generally leads to naked singularities
in the supergravity solution \cite{cells,ypqdef,kt,pt}.

It is worth mentioning that the group of automorphisms of this group
is not $SL(2,\mathbb{Z})$.  To see this, consider a general automorphism
of the type
\bea
\label{auto}
&&A\mapsto A'= A^{a_1}B^{b_1}C^{c_1}D^{d_1}, \qquad
B\mapsto B'= A^{a_2}B^{b_2}C^{c_2}D^{d_2}, \nonumber \\
&&C\mapsto C'= A^{a_3}B^{b_3}C^{c_3}D^{d_3}, \qquad 
D\mapsto D'= A^{a_4}B^{b_4}C^{c_4}D^{d_4},
\label{eq:auto}
\eea
which maps $A$, $B$, $C$ and $D$ to {\it arbitrary} elements of the group.  
In order to investigate the primed elements, we first work out the general
commutation relation between two group elements.  The result is simply
\bea
(A^{a_i}B^{b_i}C^{c_i}D^{d_i})(A^{a_j}B^{b_j}C^{c_j}D^{d_j})&&\nonumber\\
&&\kern-9em=(A^{a_j}B^{b_j}C^{c_j}D^{d_j})(A^{a_i}B^{b_i}C^{c_i}D^{d_i})
C^{a_ic_j-a_jc_i}D^{a_ic_j-a_jc_i+b_i\frac{a_j(a_j+1)}{2}
-b_j\frac{a_i(a_i+1)}{2}}.\nonumber\\
\eea
We further denote the order $A$ as $p_a$ (so that $A^{p_a}=1$), and so on for
the other operators.  

To show that (\ref{eq:auto}) is an automorphism, we need to demonstrate that
the primed elements satisfy the same commutation relations as the original
elements.  We start with the $A'$ and $B'$ relation.  Demanding that
$A'B'=B'A'C'$ implies that the $C'$ element must be of the form
\be
C'\equiv C^{a_1b_2-a_2b_1}D^{a_1c_2-a_2c_1+b_1\frac{a_2(a_2+1)}{2}
-b_2\frac{a_1(a_1+1)}{2}}.
\label{eq:cpdef}
\ee
Hence in the general form of (\ref{auto}) one must require $a_3=b_3=0$.  
Next, by turning to the commutation relation $A'C'=C'A'D'$, we find that
$D'$ is
\be
D'\equiv D^{a_1 c_3}=D^{a_1 (a_1b_2-a_2b_1)},
\label{eq:dpdef}
\ee
where we used the relation $c_3=a_1b_2-a_2b_1$ implicit in (\ref{eq:cpdef}).
This demonstrates that $D'$ automatically commutes with everything, and so
the rest of its commutation relations are automatically satisfied.

Finally, however, for (\ref{eq:auto}) to be an automorphism, the $B'$ and
$C'$ elements must commute.  We now note the problem: $B'$ contains $A$ and
$C'$ contains $C$, so there is an obstruction to their commutation.  From
(\ref{eq:cpdef}) and (\ref{eq:dpdef}), we find simply
\be
B'C'=C'B'D^{a_2(a_1b_2-a_2b_1)}.
\ee
Demanding that the right hand side is equal to $C'B'$ gives rise to
the condition
\be
a_2(a_1b_2-a_2b_1)\equiv 0 \; (\mbox{mod} \; p_d).
\label{eq:dcond}
\ee
This is the only condition that needs to be met.  All other commutators are
trivially satisfied.

The condition (\ref{eq:dcond}), however, is enough to indicate that the
automorphism (\ref{eq:auto}) cannot be identified with the $SL(2,\mathbb{Z})$
that acts in the conformal case where $D\equiv 1$.  (For $D\equiv1$, we may
formally take $p_d=1$, in which case the above condition becomes trivial.)
One way to see this is that the original $SL(2,\mathbb{Z})$ acts with
$a_1b_2-a_2b_1=1$, whereupon (\ref{eq:dcond}) reduces to $a_2\equiv0
\;(\mbox{mod}\;p_d)$.  This is a restriction of $SL(2,\mathbb{Z})$
transformations to the shifts $\tau\rightarrow\tau+1$ only.

We interpret this breaking of the $SL(2,\mathbb{Z})$ symmetry simply
as a reflection of the fact that the background breaks the symmetry.
Namely, the background contains only D5 branes and no NS5 branes.
Hence an $SL(2,\mathbb{Z})$ transformation of the form $\tau\rightarrow-1/\tau$
interchanging NS5 with D5 cannot act as an automorphism of the quiver
theory.  On the other hand, it is interesting to consider the
$SL(2,\mathbb{Z})$ action on the group elements
\bea
&&A\mapsto A'=A^{a_1}B^{b_1},\qquad B\mapsto B'=A^{a_2}B^{b_2},\nonumber\\
&&C\mapsto C'=C,\qquad\qquad D\mapsto D'=D,
\eea
with $a_1b_2-a_2b_1=1$.  The resulting commutation relations on the
primed elements take the form
\be
A'B'=B'A'C',\qquad
A'C'=C'A'D'^{a_1},\qquad  B'C'=C'B'D'^{a_2}
\label{eq:sdual}
\ee
(with $D'$ a central element).  This is suggestive of the $A'C'$
non-commutativity being related to fractional D5 branes, and the $B'C'$
non-commutativity being related to fractional NS5 branes.  It would, of
course, be interesting to find a more symmetric description of this
background.  Our guess, implicit in (\ref{eq:sdual}), is that when
NS5 branes wrapping two-cycles are included the group is further extended
by an operator that reflects the presences of background NS5 flux.

%%%%%%%%%%%%%%%%%%%%%%%%%%%%%%%%%%%%%%%%%%%%%%%%%%%%%%%%%%%%%%%%%%%%%%%%%%%
\section{Conclusions}\label{conclusions}
%%%%%%%%%%%%%%%%%%%%%%%%%%%%%%%%%%%%%%%%%%%%%%%%%%%%%%%%%%%%%%%%%%%%%%%%%%
In this paper we have established that  a large class of nonconformal quiver gauge theories
admits the action of a finite group which is the central extension of the finite Heisenberg group
that acts on the conformal case.

The existence of a finite Heisenberg group for a large class of quiver gauge theories was established explicitly in
\cite{heisAlg}. Here we study a natural generalization to the nonconformal situation. The nonconformal quiver
gauge theories have running beta functions and are, therefore more dynamical than their conformal counterparts.
The existence of a finite group even in this dynamical case leads us to believe, based on the AdS/CFT correspondence,
 that in spaces with torsion and RR flux the operators counting the number of wrapped D-branes do  not commute and
essentially satisfy a centrally extended finite Heisenberg group.

As is well known, the nonconformal quiver gauge theories have running gauge couplings and a natural way to interpret their
RG flow is via a cascade of Seiberg dualities \cite{ks}. Towards the end of the cascade some nonperturbative superpotential
develops, along the lines of the Affleck-Dine-Seiberg  superpotential \cite{ads}. In this note we have established the existence of a
central extension of a finite Heisenberg group acting on the theory only in the ultraviolet  regime where
$N\gg M$ and thus the theory is still close to the conformal point. As discussed in various articles in the literature, for most
of the nonconformal quiver gauge theories that we discussed after a duality cascade the ranks of the gauge groups are changed
as $N\rightarrow N-M$ \cite{cells,chaos}, that is, the quiver is self similar.  The self-similarity implies that the same
algebraic structure is present with the appropriate redefinitions of the roots $\omega_i$'s of section \ref{meat}
to incorporate $N\rightarrow N-M$. It would be really interesting to follow the action of the
group deep into the cascade where nonperturbative effects become important, that is assuming that
$M$ is a factor of $N$ as in $M=pN$.  After approximately $p$ steps the
structure of the theory changes significantly. We hope to return to this interesting question in the future.

%%%%%%%%%%%%%%%%%%%%%%%%%%%%%%%%%%%%%%%%%%%%%%%%%%%%%%%%%%%%%%%%%%%%%%%%%%%
\section*{Acknowledgments}
%%%%%%%%%%%%%%%%%%%%%%%%%%%%%%%%%%%%%%%%%%%%%%%%%%%%%%%%%%%%%%%%%%%%%%%%%%%
We are grateful to D. Belov, S. Benvenuti, J. Davis and  P. de Medeiros for comments and correspondence.
This work  is  partially supported by Department of Energy under grant
DE-FG02-95ER40899 to the University of Michigan

%%%%%%%%%%%%%%%%%%%%%%%%%%%%%%%%%%%%%%%%

\end{document}